\documentstyle[preprint,eqsecnum,aps]{revtex}

\begin{document}


\draft 
\preprint{}
\topskip 2cm

\begin{center}
{\bf DIMENSIONALLY CONTINUED OPPENHEIMER-SNYDER\\ 
GRAVITATIONAL COLLAPSE. I -- SOLUTIONS IN EVEN DIMENSIONS} \\
\vskip 2cm
{\bf Anderson Ilha} \\
\vskip 0.3cm
{\scriptsize  Departamento de Astrof\'{\i}sica,
	      Observat\' orio Nacional-CNPq,} \\
{\scriptsize  Rua General Jos\'e Cristino 77,
	      20921 Rio de Janeiro, Brasil,} \\
\vskip 0.6cm
{\bf Jos\'e P. S. Lemos} \\
\vskip 0.3cm
{\scriptsize  Departamento de Astrof\'{\i}sica,
	      Observat\' orio Nacional-CNPq,} \\
{\scriptsize  Rua General Jos\'e Cristino 77,
	      20921 Rio de Janeiro, Brasil,} \\
{\scriptsize  \&} \\
{\scriptsize  Departamento de F\'{\i}sica,
	      Instituto Superior T\'ecnico,} \\
{\scriptsize  Av. Rovisco Pais 1, 1096 Lisboa, Portugal.} \\

\end{center}

\title{Dimensionally Continued Oppenheimer-Snyder\\ 
Gravitational Collapse. I -- Solutions in Even Dimensions}
\author{Anderson Ilha}
\address{Departamento de Astrof\'{\i}sica,
	      Observat\' orio Nacional-CNPq, \\
              Rua General Jos\'e Cristino 77,
	      20921 Rio de Janeiro, Brasil,}

\author{Jos\'e P. S. Lemos}
\address{Departamento de Astrof\'{\i}sica,
	      Observat\' orio Nacional-CNPq, \\
              Rua General Jos\'e Cristino 77,
	      20921 Rio de Janeiro, Brasil,  \\  \& \\
              Departamento de F\'{\i}sica,
	      Instituto Superior T\'ecnico, \\
              Av. Rovisco Pais 1, 1096 Lisboa, Portugal,}

\vfill\eject

\begin{abstract}
The extension of the general relativity theory to higher dimensions, so
that the field equations for the metric remain of second order, is done
through the Lovelock action. This action can also be interpreted as the
dimensionally continued Euler characteristics of lower dimensions. The
theory has many constant coefficients apparently without any physical
meaning. However, it is possible, in a natural way,  to reduce to two
(the cosmological and Newton's constant) these several arbitrary
coefficients, yielding a restricted Lovelock gravity. In this process
one separates theories in even dimensions from theories in odd
dimensions.  These theories have static black hole solutions.  In
general relativity, black holes appear as the final state of
gravitational collapse.  In this work, gravitational collapse  of a
regular dust fluid in even dimensional restricted Lovelock gravity is
studied.  It is found that black holes emerge as the final state for
these regular initial conditions.

\end{abstract}
\bigskip

\widetext

\section{Introduction}
\label{sec:level1}

The theory of general relativity has predicted the existence of black holes 
in our universe. The relevance of black holes comes from their unavoidable 
emergence in complete gravitational collapse of astrophysical objects, as 
was first shown by Oppenheimer and Snyder \cite{Oppie}. 

General relativity was formulated in a spacetime 
with four dimensions, of course. However, there are theoretical hints that 
we might live in a world with more dimensions. Kaluza-Klein theories,  
with the help of extra compact dimensions, try to unify the gravitational 
field to the other known gauge fields. String theory, a theory that 
embraces gravity and other interactions in a unified picture, lives, in 
its heterotic form, in ten dimensions. Turning-off all other fields, 
the low-energy limit of string theory yields Einstein-Hilbert action plus 
terms involving quadratic powers in the curvature. 
For the quantum version of the theory, these quadratic corrections 
should be proportional to the Gauss-Bonnet term in order to give 
ghost-free meaningful interactions. 

On the other hand, one can ask what is the most natural generalization 
of Einstein gravity to other dimensions while keeping the same degrees 
of freedom. Although pure general relativity can be formulated in other 
dimensions, when one goes to dimensions higher than four it is not anymore 
unique. The natural generalization is given by the Lovelock action  
\cite{Lovie} so that the field equations for the metric remain of 
second order. 
The theory can also be considered as the topological extension of 
Eintein-Hilbert action \cite{TZ}. In this theory new terms make their 
appearance by taking into the action the Euler densities of the spaces 
with dimensions lower than the space in consideration. The Euler density 
of the space in consideration yields a topological term only, with no 
pure dynamical content. In four dimensions 
one has to take in consideration two Euler densities. The Euler density 
of the 0-dimensional space which is proportional to $\sqrt{-g}$, 
and the Euler density of the 2-dimensional space, proportional to 
$\sqrt{-g} R$, where $g$ is the determinant of the metric and $R$ the 
Ricci curvature scalar. Thus Lovelock gravity in four dimensions reduces 
to Einstein gravity, with action 
$\frac{1}{16\pi G}\int d^4 x \sqrt{-g}(-2\Lambda+R)$, 
where $\Lambda$ and $G$ are the cosmological and Newton's constant, 
repectively. A similar construction and 
action is obtained for three dimensions. In six dimensions one 
has still to add the Euler characteristic of four dimensional space, 
i.e. the Gauss-Bonnet term, to have the Lanczos action, given by,  
$\frac{1}{16\pi G}\int d^6 x\sqrt{-g}\left(-2\Lambda+R+
\alpha_2(R_{\alpha\beta\gamma\sigma}R^{\alpha\beta\gamma\sigma}
-4R_{\alpha\beta}R^{\alpha\beta}+R^2)\right)$, where $\alpha_2$ 
is a new constant. For each two new dimensions there exists a 
new constant $\alpha_p$. 
These constants do not seem to have a direct physical meaning.  In 
order to find a meaningful set of constants in any dimension ${\cal D}$, 
it was proposed in \cite{BTZ} a method which 
restricts drasticaly the number of independent constants to two, 
$G$ and $\Lambda$, thus yielding a restricted Lovelock gravity. 
This method separates, in a natural manner, 
theories in even dimensions (${\cal D}=2n$, with $n=1,2,..$) 
from theories in odd dimensions  (${\cal D}=2n+1$).

Several static and cosmological solutions within Lovelock gravity have
been found \cite{several}. In the restricted setting of \cite{BTZ},
where the independent constants are reduced to two, wormhole \cite{Li}
and black hole solutions have also been found \cite{BTZ} both in even
and odd dimensions. Higher dimensional black holes may shed some light 
on the understanding of non-perturbative effects in quantum
gravity. They can also expose which of the features of the usual 
four-dimensional black hole solutions remain in higher dimensions.  
Since in general relativity black holes appear as the final state of
gravitational collapse it is important to know if the black hole
solutions found in Lovelock gravity can, in an analogous manner, form
from gravitational collapse. We will show that, indeed, black holes
form from regular initial data. 
A possible scenario for the occurrence of this collapse in ${\cal
D}$ dimensions, would be in the very early universe, before the ${\cal
D}-4$ extra dimensions have been compactified. In turn, these newly
formed higher dimensional black holes could play a role in the
compactification process.

In this work we study gravitational collapse of a dust fluid in 
Lovelock gravity within the context of the restricted coefficients 
found in \cite{BTZ}. We will analyse the even dimensional case only.  
For the odd dimensional case see \cite{ilhalemos2}. We thus generalyze  
the Oppenheimer-Snyder collapse. In section II the  Lovelock
gravity for restricted coefficients is presented. 
In section III we display the static solutions in even dimensions 
found in \cite{BTZ}. In section IV we find some cosmological 
solutions for perfect fluids. In section V we match the solutions found 
in section IV to the solutions of section III. Finally, in section 
VI we show that black holes can form through gravitational collapse 
in Lovelock gravity. Section VI presents some conclusions. In the rest 
of the paper we usually do $G=c=1$.

\section{The Lovelock theory}

The most general action in ${\cal D} \geq 3$ spacetime dimensions that
yields    the same degrees of freedom of Einstein's theory is the so
called Lovelock action, given by \cite{Lovie,TZ}
\begin{equation}
S = \int\, {\cal L}_{\cal D} =  \kappa\,\sum_{p=0}^{\left[({\cal D}-1)/2\right]}\,
\alpha_{p}\,\int_{M}\,
\epsilon_{a_{1} \cdots
a_{{\cal D}}}\,R^{a_{1}a_{2}}\,\wedge\,\cdots\,
\wedge\,R^{a_{2p-1}a_{2p}}\,
\wedge\,e^{a_{2p+1}}\,\wedge\,\cdots\,\wedge\,
e^{a_{{\cal D}}} + S_{m}\, ,                                                        \label{2.1}
\end{equation}
where $R^{ab}={d}\omega^{ab}+
\omega^{a}_{\,\,c}\,\wedge\,\omega^{cb}$ is the curvature two-form,
$e^{a}$ is the local frame one-form,  and $\omega^{ab}$ is the spin
connection, with $a_{i} = {0, 1, \dots, {\cal D} -1}$.  The symbol $[
]$ over the summation symbol means one should take the integer part of
$({\cal D}-1)/2$. $S_{m}$ is a phenomenological action wich describes
the macroscopic matter sources.

In general, the constant coefficients $\alpha_{p}$ are arbitrary.
However, it is shown in \cite{BTZ} that taking certain special choices 
one is able to get simple meaningful solutions. Following \cite{BTZ} one
first considers embedding the Lorentz group $SO({\cal D}-1,1)$ into de
anti-de Sitter group $SO({\cal D}-1,2)$, and then separates into two
distinct classes of Lagrangians: Lagrangians for even dimensions and 
Lagrangians for odd dimensions.  

For even dimensions, ${\cal D} = 2n, \;\; (n = 2, 3, \dots)$, 
one chooses the following Lagrangian
\begin{equation}
{\cal L}_{2n}  = \kappa\,\hat{R}^{A_{1}A_{2}}\,\wedge\,
\hat{R}^{A_{3}A_{4}}\,\wedge\,\cdots\,
\wedge\,
\hat{R}^{A_{{\cal D}-1}A_{{\cal D}}}\,Q_{A_{1}A_{2} \cdots A_{{\cal D}}}
\label{2.2}
\end{equation}
with $A_{1}, A_{2} = 0, 1, \dots , {\cal D}$ being anti-de Sitter
indices. $\hat{R}^{A_{1}A_{2}}$ is the anti-de Sitter curvature
two-form constructed with the $SO({\cal D}-1,2)$ connection,
$W_{A_{1}A_{2}}$.  In order to yield a non-trivial action, 
the tensor $Q$ is chosen to  be an invariant tensor under the Lorentz 
group only, i.e., 
$Q_{A_{1}A_{2} \cdots A_{\cal D}} = 
\epsilon_{ a_{1} a_{2} \cdots a_{\cal D} }$ for $A_{i} = a_{i} \;\; 
(i = 0, \dots \, {\cal D} - 1)$ and zero otherwise.  
Decomposing the connection $W^{AB}$ into the connection under ${\cal
D}$ rotations, $\omega^{ab}$, and inner translations, $e^{a}$, one
finds the anti-de Sitter curvature $\hat{R}$ in terms of the Lorentz
curvature $R$
\begin{equation}
\hat{R}^{ab} = R^{ab} + \frac{1}{l^2}\, e^{a}\,\wedge\,e^{b}, 
\label{2.3}
\end{equation}
where $l$ is a scale factor which is to be related to the cosmological
constant $l^2=-\frac{1}{\Lambda}$. 
Using (\ref{2.3}) one finds that the Lagrangian (\ref{2.2})
can be put in the form
\begin{equation}
{\cal L}_{2n}  = \kappa\,\left(R^{a_{1}a_{2}} + 
\frac{1}{l^2}\,e^{a_{1}}\,\wedge\,e^{a_{2}}\right)
\,\wedge\,
\cdots\,\wedge\,\left(R^{a_{{\cal D}-1}a_{{\cal D}}} + 
\frac{1}{l^2}\,e^{a_{{\cal D}-1}}\,\wedge\,
e^{a_{\cal D}}\right)\,\epsilon_{a_{1}a_{2}\cdots 
a_{{\cal D}}}.
\label{2.4}
\end{equation}
This Lagrangian gives the Born-Infeld gravity \cite{giambiagi}.
Comparison of (\ref{2.4}) and (\ref{2.1}) gives the 
coefficients $\alpha_{p}$, 
\begin{equation}
\alpha_{p} = \kappa\,\left(\stackrel{n}{{}_p}\right)\,
l^{-{\cal D} + 2p}, \;\;\; 
{\cal D} = 2n
\label{2.5}
\end{equation}
where, for convenience one can choose $\kappa$ as
\begin{equation}
\kappa = \frac{l^{{\cal D}-2}}{32\,\pi\,G\,n}, 
\;\;\; {\cal D}=2n 
\label{2.6}
\end{equation}

For odd dimensions, ${\cal D} = 2n -1$, one can find a construction
similar to the Chern-Simons action construction in three dimensions.
One starts with the Euler density in one dimension above ${\cal D}$,
$E_{2n}$, which is an exact form, and can be written as an
exterior derivative of a Lagrangian in $2n-1$ dimensions, i.e.,
$E_{2n} = {d}{\cal L}_{2n - 1}$, see \cite{BTZ}. Since 
gravitational collapse in odd dimensions has different features 
from collapse in even dimensions we study odd dimensional collapse 
in another work \cite{ilhalemos2}.

Given the action (\ref{2.1}), the field equations are obtained by the
variation with respect to the one-forms $e^{a}$.  Under the assumption
of zero torsion, the variation with respect to the spin connection
$\omega^{ab}$ vanishes identically. Although the equations have powers
in the curvatures, they remain by construction  second order in the
metric. The field equations are given by
\begin{equation}
-\kappa\,\sum_{p=0}^{\left[({\cal D}-1)/2\right]}\,
\alpha_{p}\,({\cal D} - 2p)\,\epsilon_{a_{1} ..
a_{{\cal D}}}\,R^{a_{1}a_{2}}\,\wedge\,..\,\wedge\,
R^{a_{2p-1}a_{2p}}\,
\wedge\,e^{a_{2p+1}}\,\wedge\,..\,\wedge\,
e^{a_{{\cal D}-1}} = Q_{a_{{\cal D}}}\, ,
\label{2.7}
\end{equation}
where $Q_{a_{{\cal D}}}$ is a $({\cal D} -1)$-form
associated with the energy momentum tensor $T^{a}_{b}$ through the following expression
\begin{equation}
Q_{i} = \frac{1}{({\cal D} -1)!}\,T^{a_{1}}_{i}\,
\epsilon_{a_{1} \cdots a_{{\cal D}}}\,e^{a_{2}}\,\wedge\,\cdots\,\wedge\,
e^{a_{{\cal D}}}\, .
\label{2.8}
\end{equation}

\section{Exterior Vacuum Solutions}

In the vacuum  all components of the energy-momentum tensor vanish, so that 
the field equations (\ref{2.7}) are given by
\begin{equation}
-\kappa\,\sum_{p=0}^{n}\,\alpha_{p}\,({\cal D} - 2p)\,
\epsilon_{a_{1} \cdots
a_{{\cal D}}}\,R^{a_{1}a_{2}}\,\wedge\,\cdots\,\wedge\,
R^{a_{2p-1}a_{2p}}\,
\wedge\,e^{a_{2p+1}}\,\wedge\,\cdots\,\wedge\,
e^{a_{{\cal D}-1}} =  0\, .
\label{3.1}
\end{equation}
Inserting the coeficients  $\alpha_{p}$ and the constant $\kappa$ given
in (\ref{2.5}) and  (\ref{2.6}) in equation (\ref{3.1}), one gets 
for even dimensions (${\cal D}=2n$), 
\begin{equation}
(R^{a_1 a_2} + l^{-2}\, e^{a_1}\, \wedge \,
e^{a_2})\, \wedge\, .. \, \wedge\,
(R^{a_{2n-3} a_{2n-2}} + l^{-2}\, e^{a_{2n-3}}\,
\wedge\, e^{a_{2n-2}})\, \wedge\,
e^{a_{2n-1}}\,\epsilon_{a_1 a_2..a_{2n}} = 0.
	\label{3.2}
\end{equation}
We consider now a static, spherical symmetric spacetime. 
One can write the metric in the following form, 
\begin{equation}
{d}s^2_{+}=-g^2(r_{+})\, 
{d}t_{+}^2 + g^{-2}(r_{+})\,{d}r_{+}^2 + 
r_{+}^2\,{d}\Omega_{{\cal D}-2}^2\, ,
\label{3.3}
\end{equation}
where $t$ and $r$ are the time and radial coordinates and
${d}\Omega_{{\cal D}-2}^2$ is the arc-element of a unit (${\cal
D}-2$)-sphere.  The subscript + reminds that (\ref{3.3}) is to be
viewed  as an exterior solution.  With  metric (\ref{3.3}) and
equations (\ref{3.1}) and (\ref{3.2}), Ba\~nados, Teitelboim and
Zanelli found the following exact solution for ${\cal D}=2n$
\cite{BTZ},
\begin{equation}
{d}s^2_{+} = -\left[1 - (2M/r_{+})^{\frac{1}{n - 1}} +
 (r_{+}/l)^2\right]\,{d}t_{+}^2 +
\frac{{d}r_{+}^2}{1 - (2M/r_{+})^{\frac{1}{n - 1}} + (r_{+}/l)^2} + r_{+}^2\,
 {d}\Omega^2_{{\cal D}-2}. 
\label{3.4}
\end{equation}
These solutions describe black holes. We will show that they also
represent the exterior vacuum solution to a collapsing (or expanding) 
dust cloud in Lovelock's theory.

\section{Interior Matter Solutions}
	
The interior spacetime is modeled by a homogeneous 
colapsing (or expanding) dust cloud, whose metric is described by the
Friedmann-Robertson-Walker in ${\cal D}$ dimensions
\begin{equation}
ds^{2} = -dt^{2} + a^{2}\left(t\right)\,
 \left[\frac{dr^{2}}{1 - k\,r^{2}} + 
r^{2}\,d\Omega^{2}_{{\cal D}-2} \right]\, .
	\label{4.1}
\end{equation}
The coordinates $t$ and $r$ are comoving coordinates 
(we omit throughout the subscript $-$ to indicate an interior solution). 
The constant $k$ can take the values, $k = 0, \pm 1$. 
The energy-momentum tensor for a perfect fluid is given by
\begin{equation}
T_{\alpha\beta} = (\rho+p)\,u_{\alpha}\,u_{\beta} + p g_{\alpha\beta}\, ,
\label{4.2}
\end{equation}
where $\rho$ is the energy-density, $p$ the pressure, 
and $u^{\alpha}$ is the ${\cal D}$-velocity of the  fluid. 
From (\ref{4.1})-(\ref{4.2}) and 
Lovelock equations (\ref{2.7}) we obtain
\begin{equation}
({\cal D}-1)!\,\sum_{p}\,\alpha_{p}\,({\cal D}-2p)\,\left(\frac{k + \dot{a}^{2}}{a^{2}}\right)^{p} = \rho + p
\label{4.3}
\end{equation}
and
\begin{equation}
({\cal D}-2)!\,\sum_{p}\,({\cal D}-2p)\,\left(\frac{k + \dot{a}^{2}}{a^{2}}\right)^{p-1}\,
\left(2p\,\frac{\ddot{a}}{a} - ({\cal D} - 2p - 1)\,\frac{k + \dot{a}^{2}}{a^{2}}\right) = 0\, ,
\label{4.4}
\end{equation}
where the coefficients $\alpha_{p}$ are given in (\ref{2.5}), and
$\kappa$ is given in (\ref{2.6}). Rearranging (\ref{4.3})-(\ref{4.4}) 
we obtain the following system of equations 
\begin{equation}
-B\,\frac{{d}}{{d}\tau}\left(\frac{\dot{a}}{a}\right) 
+ \frac{k}{a^2} = \rho + p
\label{4.5}
\end{equation}
\begin{equation}
({\cal D} - 1)\, B\,\left(\frac{\dot{a}}{a}\right)\left[-\frac{k}{a^2} +
\frac{{d}}{{d}\tau}\left(\frac{\dot{a}}{a}\right)\right] = 
\dot{\rho}
\label{4.6}
\end{equation}
where, 
\begin{equation}
B \equiv ({\cal D}-2)!\,\sum_{p}\,\alpha_{p}\,2p\,({\cal D}-2p)\,
\left(\frac{\dot{a}^{2}+k}{a^2}\right)^{p-1}\, .
\label{4.7}
\end{equation}

Equations (\ref{4.5})-(\ref{4.6}) have a first integral given by
\begin{equation}
\dot{a}^2 = -k - \left(\frac{a}{l}\right)^2 + \left(\frac{a}{l}\right)^2 
\left[ \frac{16\,\pi\,l^2\,
\rho_{0}}{({\cal D}-1)!}\,\left(\frac{a_{0}}{a}\right)^{{\cal D} - 
1}\right]^{2/({\cal D} -2)}\, , 
\label{4.8}
\end{equation} 
where $\rho_0$ and $a_0$ are constants.

The Ricci quadratic scalar and the Kretschmann scalar are given by
\begin{equation}
R^{ab}R_{ab} = 
-({\cal D} - 1)^{2}\,\left(\frac{\ddot{a}}{a}\right)^{2} + 
({\cal D} - 1)\,
\left[\frac{\ddot{a}}{a} + ({\cal D} - 2)\,\frac{\dot{a}^{2} + 
k}{a^{2}}\right]^{2}\, , 
\label{4.9}  
\end{equation}
\begin{equation}
R^{abcd}R_{abcd} = ({\cal D} - 1)\,
\left[\left(\frac{\ddot{a}}{a}\right)^{2} + 
\left(\frac{\dot{a}^{2} + k}{a^{2}}\right)^{2}\right] 
\label{4.10}
\end{equation}
respectively.

Taking  (\ref{4.5}) and  (\ref{4.6}) yields 
\begin{equation}
\dot{\rho} + ({\cal D} - 1)\,(\rho + p)\,\frac{\dot{a}}{a} = 0\, . 
\label{4.11}
\end{equation}
We now assume a dust fluid, $p=0$. For such an equation of 
state we can integrate  (\ref{4.11}) to give 
\begin{equation}
\rho = \rho_{0}\,\left(\frac{a_{0}}{a}\right)^{{\cal D}-1}\, , 
\label{4.12}
\end{equation}
where $\rho_0$ and $a_0$ are the constants defined above.

In general it is not possible to obtain an exact analytical solution of
(\ref{4.8}) for $k = \pm 1$. However, restricting to ${\cal D} =4$, 
one of course obtains the Lemaitre models, of which the closed and open
Friedmann universes are the particular cases found for 
$l\rightarrow \infty$. 
For $k = -1$ and ${\cal D} \neq 4$ there is a
special solution with zero matter content, taken in the limit $l
\rightarrow \infty$ and given by
\begin{equation}
a\left(t\right) = \pm t
\label{4.13}
\end{equation}
where for the --- sign one takes $-\infty < t < 0$ and for the + sign 
$0 < t < \infty$. There are no singularities in these solutions since, 
as one can show, the curvature scalars  are null. This solution indicates 
that the higher order terms in the curvature appearing in the Lovelock 
gravity act, in some sense, as matter terms \cite{far}.

The marginally bound case, $k = 0$, allows a second integral of 
(\ref{4.8}) given by 
\begin{equation}
a = a_{0}\,\left\{\frac{16\,\pi}{({\cal 
D}-1)!}\,\rho_{0}\,l^2\,
\sin^{{\cal D}-2}\left[-\frac{{\cal D} - 1}{{\cal D} - 2}\,\left(\frac{t - 
t_{0}}{l}\right)\right]\right\}^{\frac{1}{{\cal D}-1}}\, .
\label{4.14}
\end{equation}
where $t_{0}$ gives the time for which $a = 0$, and without 
loss of generality, one can put $t_{0} = 0$. We take 
$-\pi < t <0$. For $ -\pi < t < \pi / 2$ the cloud 
is expanding. For $-\pi / 2 < t < 0$ the cloud is collapsing. 
And $t={\pi\over2}$ is a moment of time-symmetry. 

Inserting (\ref{4.14}) in (\ref{4.12}) we obtain the evolution of the 
density in the $k=0$ dust model,  
\begin{equation}
\rho(t) = \frac{({\cal D}-1)!}{16\,\pi}
\left(\frac{1}{l}\right)^{2}\,\sin^{-({\cal D}-2)}
\left[-\frac{{\cal D}-1}{{\cal D}-2}\,
\left(\frac{t}{l}\right)\right]\, .
\label{4.15}
\end{equation}

The curvature scalars (\ref{4.9})-(\ref{4.10}) and the density 
(\ref{4.15}) diverge, 
at $t= -\pi$ (representing the appearance of a singularity), 
and $t = 0$ (denoting the formation of a singularity).

We can now take the limit of zero cosmological constant. Indeed, 
expanding (\ref{4.14}) in powers of $(1/l)$ yields 
\begin{equation}
\left(\frac{a}{a_{0}}\right)^{{\cal D}-1} 
\approx \alpha
\left[     
\left(\,\overline{t}\,\right)^{{\cal D}-2}\,
\left(\frac{1}{l}\right)^{{\cal D}-4} - 
\frac{{\cal D}-2}{3}
\left(\,\overline{t}\,\right)^{{\cal D}}\,
\left(\frac{1}{l}\right)^{{\cal D}-2} 
+ {\cal O}^{{\cal 
D}}\left(\frac{1}{l}\right)
\right]\, .
\label{4.16}
\end{equation}
where $\alpha\equiv \frac{16\,\pi}{({\cal D}-1)!}\,\rho_{0}$ and, 
$\overline{t}  \equiv - \frac{{\cal D} - 1}{{\cal D} - 2}\,  {t}$. 

When $D\ne4$ and $l \rightarrow \infty$ one gets $a=0$, i.e, 
no physical solution. 
For ${\cal D} = 4$ and $l \rightarrow \infty$ one recovers the usual 
Friedmann $k=0$ case, 
\begin{equation}
\left(\frac{a}{a_{0}}\right)^{3} \approx  \left[ \frac{3}{2}\,\sqrt{\frac{8\,\pi}{3}\,\rho_0}
\,t\right]^{2}, 
\label{4.19}
\end{equation}
whereas the density goes like $\rho \sim t^{-2}$.

\section{Junction Conditions}

Now we match the exterior and interior spacetimes found in sections III 
and IV, respectively, across an interface of separation $\Sigma$. The 
junctions conditions are \cite{Israel}  
\begin{eqnarray}
\left. ds^{2}_{+}\right]_{\Sigma} &=& \left. ds^{2}_{-}\right]_{\Sigma} \label{5.1} \\ 
\left. K_{\alpha\beta}^{+}\right]_{\Sigma} &=& \left. 
K_{\alpha\beta}^{-}\right]_{\Sigma} \label{5.2}
\end{eqnarray}
where $K_{\alpha\beta}$ is the extrinsic curvature, 
\begin{equation}
K_{\alpha\beta}^{\pm}= -n_{\epsilon}^{\pm}\,\frac{\partial^{2}
x_{\pm}^{\epsilon}}
{\partial \xi^{\alpha}\partial \xi^{\beta}} - n_{\epsilon}^{\pm}\,
\Gamma_{\gamma\delta}^{\epsilon}\,
\frac{\partial x_{\pm}^{\gamma}}{\partial \xi^{\alpha}}\,
\frac{\partial x_{\pm}^{\delta}}{\partial \xi^{\beta}}
\label{5.3}
\end{equation}
and $n_{\epsilon}^{\pm}$ are the components of the unit normal vector to
$\Sigma$ in the coordinates $x_{\pm}$, and $\xi$ represents 
the intrinsic coordinates in $\Sigma$.  
The subscripts $\pm$
represent the quantities taken in the exterior and interior
spacetimes.  Both the metrics and the extrinsic curvatures in
(\ref{5.1})-(\ref{5.2}) are evaluated at $\Sigma$.  The metric
intrinsic to $\Sigma$  is written as 
\begin{equation}
ds^2_{\Sigma} = -d\tau^2 + R^2(\tau)\,
d\Omega^2_{{\cal D}-2}\, .
\label{5.4}
\end{equation}
Where $\tau$ is the proper time on $\Sigma$ and  $d\Omega^2_{{\cal D}-2}$ denotes the line element on a ${\cal D}-2$ dimensional sphere.

Using the  junction condition (\ref{5.1}), metric (\ref{5.4}) 
and the exterior metric  (\ref{3.4}) we obtain 
\begin{equation}
r_{+} = R\left(\tau \right) \, ,
\label{5.5}
\end{equation}
and 
\begin{equation}
\left[1 - (2M/r_{+})^{\frac{1}{n - 1}} + (r_{+}/l)^2\right]\, \dot{t}_{+}^{2} -  \left[1 - (2M/r_{+})^{\frac{1}{n - 1}} + 
(r_{+}/l)^2\right]^{-1}\, \dot{r}_{+}^{2} = 1\, \, ,
\label{5.6}
\end{equation}
where ${}^\cdot\equiv \frac{d}{d\tau}$, and both equations are evaluated at
$\Sigma$. From now on, we will usually omit the subscript $\Sigma$ to 
denote evaluation at the interface.
Using (\ref{5.5}) in (\ref{5.6}) we find
\begin{equation}
\frac{dt_{+}}{d\tau} = \frac{\sqrt{\left[1 -
(2M/R)^{\frac{1}{n-1}} + 
(R/l)^2\right] + \dot{R}^{2}}}{\left[1 - 
(2M/R)^{\frac{1}{n-1}} + 
(R/l)^2\right]}\, .
\label{5.7}
\end{equation}

The unit normal to $\Sigma$ in the exterior spacetime is 
\begin{equation}
n_{\epsilon}^{+} = \left(-\frac{{d}r_{+}}{{d}\tau}, 
\frac{{d}t_{+}}{{d}\tau}, 0, \cdots, 0\right).  
\label{5.8}
\end{equation}
From (\ref{5.3}) we then get
\begin{equation}
K_{\theta\theta}^{+} = R\,\sqrt{\left[1 - \left(\frac{2M}{R}
\right)^{\frac{1}{n-1}} + 
\left(\frac{R}{l}\right)^2\right] + \dot{R}^{2}}\, .
\label{5.9}
\end{equation}
In what follows the other components of $K_{ab}^{+}$ is not neeeded.

The unit normal to $\Sigma$ in the interior spacetime is 
\begin{equation}
n_{\epsilon}^{-} = \left(0, \frac{a}{\sqrt{1-k\,r^2}}, 0, \cdots, 0\right)
\label{5.10}
\end{equation}
and from  (\ref{5.3}) we have 
\begin{equation}
K_{\theta\theta}^{-} = 
R(\tau) \, \sqrt{1 - k\,\left(r_{\Sigma}
\right)^2}\, .
\label{5.11}
\end{equation}

From the condition $K_{\theta\theta}^{+} = K_{\theta\theta}^{-}$,  
(\ref{5.9}) and (\ref{5.11}) we obtain
\begin{equation}
\dot{R}^2 + \left(\frac{R}{l}\right)^2 + k\,
\left(r_{\Sigma}\right)^2 = \left(\frac{2M}{R}
\right)^{2/({\cal D} - 2)}\, .
	\label{5.12}
\end{equation}
Multiplying equation (\ref{4.8}) by $\left(r_{\Sigma}\right)^{2}$ 
we get
\begin{equation}
\dot{R}^2 + \left(\frac{R}{l}\right)^2 + k\,
\left(r_{\Sigma}\right)^2 = 
\left(\frac{R}{l}\right)^2 \,
 \left[ \frac{16\,\pi\,l^2\,
\rho_{0}}{({\cal D}-1)!}\,\left(\frac{R_{0}}{R}
\right)^{{\cal D} - 1}
\right]^{2/({\cal D} -2)}\, . 
\label{5.13}
\end{equation}
Comparing equation (5.12) and (5.13) we have 
\begin{equation}
M = \left(\frac{1}{l}\right)^{{\cal D}-4}\, 
\frac{8\,\pi}{({\cal D}-1)!}\,\rho_{0}
\,R_{0}^{{\cal D}-1}\, , 
\label{5.14}
\end{equation}
which is the mass of the cloud expressed in terms of the constants 
given in the problem. This expression is valid for any value of 
$k$,  $k = 0,\,\pm 1$.

\section{Black Hole Formation}

In order to study black hole formation in this theory we work with the 
$k=0$ model solution found in (\ref{4.14}). The interior metric is then 
\begin{equation}
ds^{2} = -dt^{2} + a^{2}\left(t\right)\,
 \left({dr^{2}} + 
r^{2}\,d\Omega^{2}_{{\cal D}-2} \right)\, .
	\label{6.1}
\end{equation}
where for convenience we rewrite (\ref{4.14}) as 
\begin{equation}
a = \left( 
\frac{2M}{{r_\Sigma}^{{\cal D}-1}} l^{{\cal D}-2}
\,\sin^{{\cal D}-2}\left[-\frac{{\cal D}-1}{{\cal D}-2}\,
\left(\frac{t}{l}\right)\right]\right)^{\frac{1}{{\cal D}-1}}\, ,
	\label{6.2}
\end{equation}
and we have used equation (\ref{5.14}) and $R_0 = a_0 r_\Sigma\, $.

The exterior metric is given in (\ref{3.4}) and as we have shown in 
section V, it is possible to make a smooth junction between both 
spacetimes.

We assume that gravitational collapse occurs for $-\frac{\pi}{2} \leq 
t\leq0$.  The time $t=-\frac{\pi}{2}$ marks the onset of collapse. At 
this moment there are no singularities in spacetime, as the curvature 
scalars (\ref{4.9})-(\ref{4.10}) and the density (\ref{4.15}) indicate.
In fact, the singularity appears only at $t=0$, where all these 
quantities blow up.

To know whether a black hole as formed or not, one has to search for  
the appearance of an apparent horizon and an event horizon. 
The apparent horizon is defined in \cite{Eardley} to be the boundary 
of the region of trapped two-spheres in spacetime. To find this boundary 
on the interior spacetime one looks for two spheres 
$Y\equiv a(t)r=$constant 
whose outward normals are null, i.e., 
\begin{equation}
\nabla\,Y\,\cdot\,\nabla\,Y = 0\, . 
\label{6.4}
\end{equation}
Using metric (\ref{6.1}) in (\ref{6.4}) yields, 
\begin{equation}
\left(\frac{d a(t)}{d t}\right)^2 = \frac{1}{r^2}.
\label{6.5}
\end{equation}
Using (\ref{6.2}) in (\ref{6.5}) gives the evolution of the apparent 
horizon in comoving coordinates, 
\begin{equation}
r = r_\Sigma \left(
\frac{1}{2m}
\right)^{1/{\cal D}-1}\,
\frac{\sin^{1/{\cal D}-1}
\left[-\frac{{\cal D}-1}{{\cal D}-2}\left(\frac{t}{l}\right)\right]
}{\cos
\left[-\frac{{\cal D}-1}{{\cal D}-2}\left(\frac{t}{l}\right)\right]
}\, . 
\label{6.6}
\end{equation}
where $m\equiv\frac{M}{l}$. 
For ${\cal D} = 4$ and $l \rightarrow \infty$ this 
expression reduces to the usual expression for the apparent 
horizon in the Friedmann metric, 
\begin{equation}
t = - \frac{2}{3}\, \left( \frac{2M}{{r_\Sigma}^3}\right)
r^{3}\, .
\label{6.8}
\end{equation}
Now, the apparent horizon first forms at the surface $r_\Sigma$. 
Then, for $r=r_\Sigma$, equation (\ref{6.6}) gives 
the time $t$ at which the apparent horizon first forms. 
On the other hand, one should also be able to find  
the formation time of the apparent horizon on the surface $\Sigma$ 
through an equation on $\Sigma$, 
equation (\ref{5.12}). Indeed, at the junction one has 
$R=a(t) r_\Sigma$. Then from junction condition (\ref{5.12}) and equation (\ref{6.5}) we have that the apparent horizon first forms when 
\begin{equation}
R\,\left[1 + \left(\frac{R}{l}\right)^{2}\right]^{({\cal D}-2)/2} = 2M.
	\label{6.10}
\end{equation}
Using (\ref{5.5}) this also gives 
$r_+\,\left[1 + \left(\frac{r_+}{l}\right)^{2}\right]^{({\cal D}-2)/2} 
= 2M$. For Friedmann $(l \rightarrow \infty$ and ${\cal D} = 4)$ 
the above expression reduces to $r_+ = 2M$, as expected. Dividing 
equation (\ref{6.10}) by $l$ and defining $x\equiv \frac{R}{l}$ 
we get 
\begin{equation}
x\,\left[1 + x^{2}\right]^{({\cal D}-2)/2} = 2m \, ,
	\label{6.11}
\end{equation}
where $m\equiv\frac{M}{l}$ as above. Now, the time of formation of 
the apparent horizon can be found through equation 
\begin{eqnarray}
R_{AH} &=& a(t_{AH})\,r_{\Sigma} \nonumber \\
&=& \left\{ 
2M\, l^{{\cal D}-2}
\,
\sin^{{\cal D}-2}\left[-\frac{{\cal D}-1}{{\cal D}-2}\,
\left(\frac{t_{AH}}{l}\right)\right]\right\}^{1/{\cal D}-1}.
\label{6.12}
\end{eqnarray}
In terms of $x$ and $m$ (\ref{6.12}) reads
\begin{equation}
x_{AH} =\left\{2m\,\sin^{{\cal D}-2}\left[-\frac{{\cal D}-1}{{\cal D}-2}\left(\frac{t_{AH}}{l}\right)\right]\right\}^{1/({\cal D}-1)}.
	\label{6.13}
\end{equation}
Given a dimension $\cal D$ and an $m$ one can obtain 
$x$ through equation (\ref{6.11}). Then equation (\ref{6.13}) 
gives implicitly $t_{AH}$, the time of the formation of the apparent 
horizon on the surface $\Sigma$. For instance, for 
${\cal D}=6$ and $m=1$ we find $t_{AH}=-0.53\, l$. 
Putting this value back in equation (\ref{6.6}) we 
verify that everything checks.

The event horizon, being a null spherical surface, is determined through 
the null outgoing lines of metric (\ref{6.1}), i.e., 
\begin{equation}
\frac{dt}{dr} = a(t)\, .
	\label{6.14}
\end{equation}
Equation (\ref{6.14}) can be put in the following integral form, 
\begin{equation}
\frac{r}{r_{\Sigma}} =  - \frac{{\cal D}-2}{{\cal D}-1}\,
\left(\frac{1}{2m}\right)^{1/({\cal D}-1)}\,\int_{u_{0}}^{u_{1}}\,
\frac{{d}u}{\sin^{({\cal D}-2)/({\cal D}-1)}(u)} \, ,
	\label{6.15}
\end{equation}
$u \equiv - \frac{{\cal D} - 1}{{\cal D} - 2}\,  \frac{t}{l}$ and $m$ 
has been defined above. Now, the time $u_1$ is 
precisely equal to the formation time of the apparent horizon, since 
in vacuum both horizons coincide \cite{hawkingellis}. One has then 
to integrate (\ref{6.15}) to find the time $u_0$ at which the event 
horizon first forms, at $r=0$. 
This can be done numerically. For ${\cal D}=6$ and $m=1$ 
we obtain $t_0 = -\frac45 u_0l=-1.57\, l$. A plot in comoving coordinates 
$(t,r)$ shows the evolution of the apparent and event horizons. 
We do this for ${\cal D} = 4, 6,10, 26$ (see figures 1,2,3,4). 
Making a matching to the vacuum exterior spacetime one finds the usual 
Penrose diagram for gravitational collapse and formation of a black 
hole in an anti-de Sitter background, see figure 5.

\section{Conclusions}

We have analysed gravitational collapse in Lovelock gravity which is 
a natural extension of Einstein's general relativity to higher 
dimensions. It was shown that within a restricted set of Lovelock 
coefficients, gravitational collapse of a regular initial non-rotating 
dust cloud proceeds, to form event and apparent horizons, and terminates 
at a spacelike curvature singularity, in much the same way as the 
Oppenheimer-Snyder collapse. As in the case of the wormhole solutions
found in \cite{Li} and the black hole solutions found in \cite{BTZ} the 
collapsing solutions studied here show that some important features 
of classical general relativity are preserved and carried 
into Lovelock gravity. 
\vskip 1cm 
\noindent {\bf Acknowledgements -} AI acknowledges a grant from CNPq 
to complete his M.S. thesis. JPSL also thanks CNPq for a research grant.

\centerline{}
\vskip 1cm
\centerline{\bf Figure Captions}
\vskip1cm

{\bf Figure 1.} 
Oppenheimer-Snyder collapse in D=4 dimensions in an asymptotically 
anti-de Sitter spacetime. The interior dust cloud in comoving coordinates 
$(t,r)$ fills the whole diagram. The left side represents the center of the 
cloud $r=0$, the right side the surface of the cloud $\frac{r}{r_\Sigma}=1$. 
The evolution of the event horizon (dashed line) and apparent horizon 
(full line) are drawn. The singularity occurs at $t=0$. 
\vskip .5cm

{\bf Figure 2.} Dimensionally continued Oppenheimer-Snyder collapse in 
D=6 dimensions in an asymptotically anti-de Sitter spacetime. 
See subtitle of figure 1 for more detailed explanation. 
\vskip .5cm

{\bf Figure 3.} Dimensionally continued Oppenheimer-Snyder collapse in 
D=10 dimensions in an asymptotically anti-de Sitter spacetime. 
See subtitle of figure 1 for more detailed explanation. 
\vskip .5cm

{\bf Figure 4.} Dimensionally continued Oppenheimer-Snyder collapse in 
D=26 dimensions in an asymptotically anti-de Sitter spacetime. 
See subtitle of figure 1 for more detailed explanation. 
\vskip .5cm

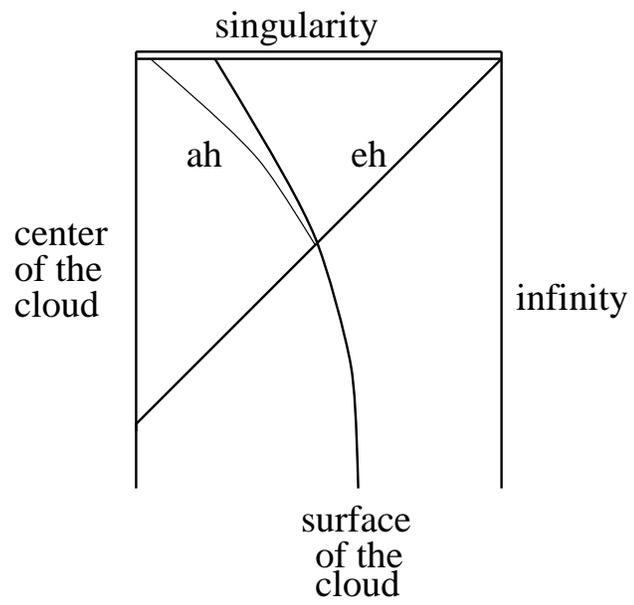
{\bf Figure 5.} Penrose diagram for the collapse of a dust cloud in 
an asymptotically anti-de Sitter spacetime. Each point in the 
diagram represents a ${\cal D}-2$ sphere. (eh=event horizon, 
ah=apparent horizon). 
\vskip .5cm


\begin{references}
\bibitem{Oppie}        J. R. Oppenheimer e H. Snyder,
                       {\it Phys. Rev.} {\bf 56}, 455 (1939).
\bibitem{Lovie}        D. Lovelock, {\it J. Math. Phys.} 
                       {\bf 12}, 498 (1971).
\bibitem{TZ}           C. Teitelboim, J. Zanelli, in {\it Constraint Theory 
                       and Relativistic Dynamics}, eds. G. Longhi, L. Lussana, 
                       (World Scientific, Singapore 1987).  
\bibitem{BTZ}          M. Ba\~nados, C. Teitelboim, J. Zanelli,
                       {\it Phys. Rev.} D {\bf 49}, 975 (1994).
\bibitem{several}      D. G. Boulware, S. Deser, {\it Phys. Rev. Lett.} 
                       {\bf 55}, 2565 (1985).
                       J.T. Wheeler, {\it Nucl. Phys.} {\bf B268}, 737 (1986);
                       R. C. Myers, J. Simon, {\it Phys. Rev. D} {\bf 38}, 
                       2434 (1988). 
                       B. Whitt, {\it Phys. Rev. D} {\bf 38}, 3001 (1988); 
                       G. A. Marug\'an, {\it Class. Quantum Grav.} {\bf 8}, 
                       935 (1991). 
\bibitem{Li}           X. Li, {\it Phys. Rev.} D {\bf 50},
                       3787 (1994).
\bibitem{ilhalemos2}   A. Ilha, J. P. S. Lemos, ``Dimensionally Continued 
                       Oppenheimer-Snyder Gravitational Collapse. 
                       II -- Solutions in Odd Dimensions'', in preparation. \bibitem{giambiagi}    M. Ba\~nados, C. Teitelboim, J. Zanelli, 
                       in {\it J. J. Giambiagi Festschrift}, 
                       edited by H. Falomir, R. Gamboa, P. Leal, 
                       F. Schasposnik (World Scientific, Singapore).  
\bibitem{far}          M. Farhoudi, gr-qc/9511047. 
\bibitem{Israel}       W. Israel, {\it Nuovo Cimento} {\bf 44B}, 463 (1967).
\bibitem{Eardley}      D. M. Eardley and L. Smarr, {\it Phys. Rev. D}  
                       {\bf 19}, 2239 (1979).
\bibitem{hawkingellis} S. W. Hawking, G. F. R. Ellis, {\it The Large 
                       Scale Structure of Space and Time}, 
                       (Cambridge University Press, Cambridge 1973).  


\end{references}
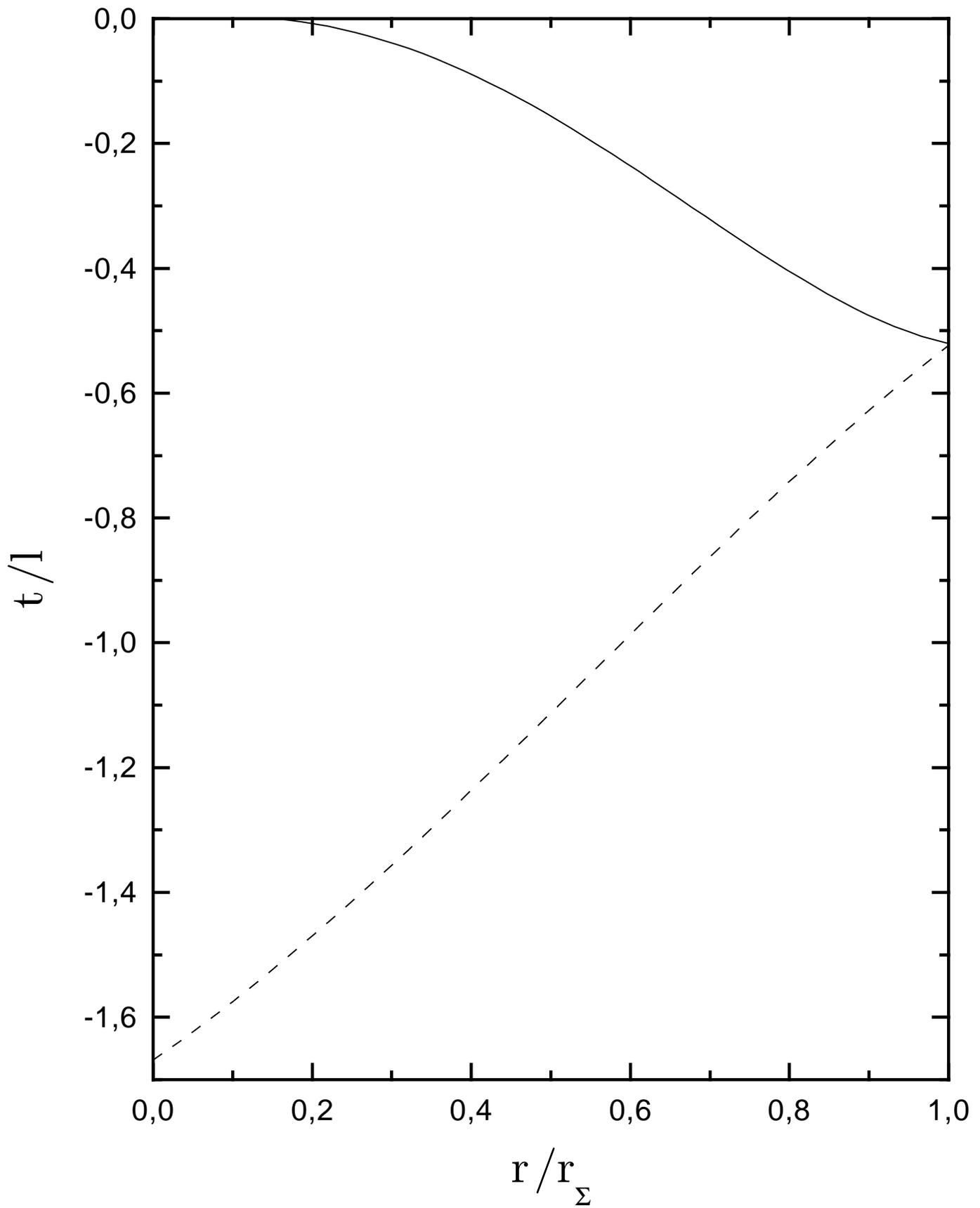
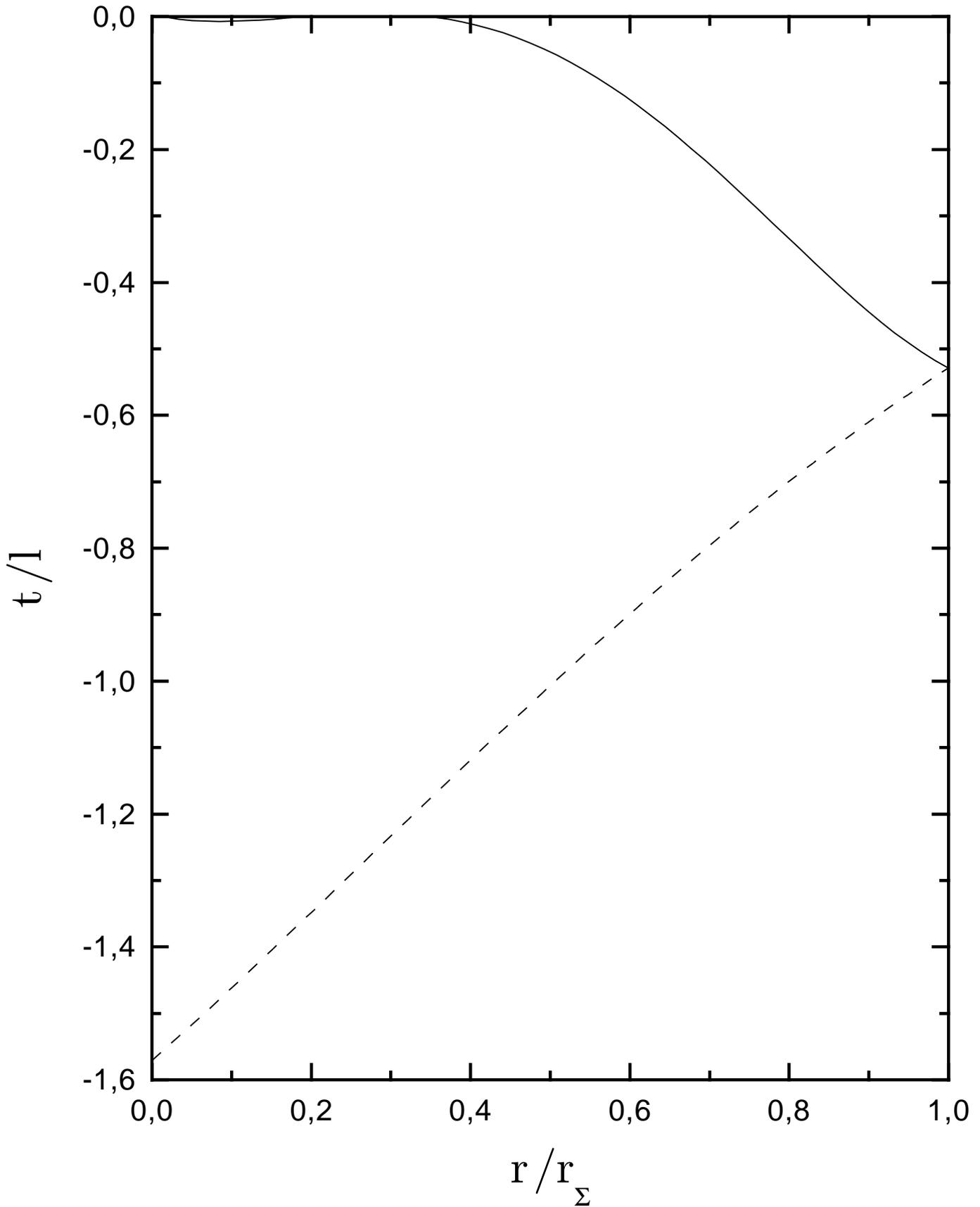
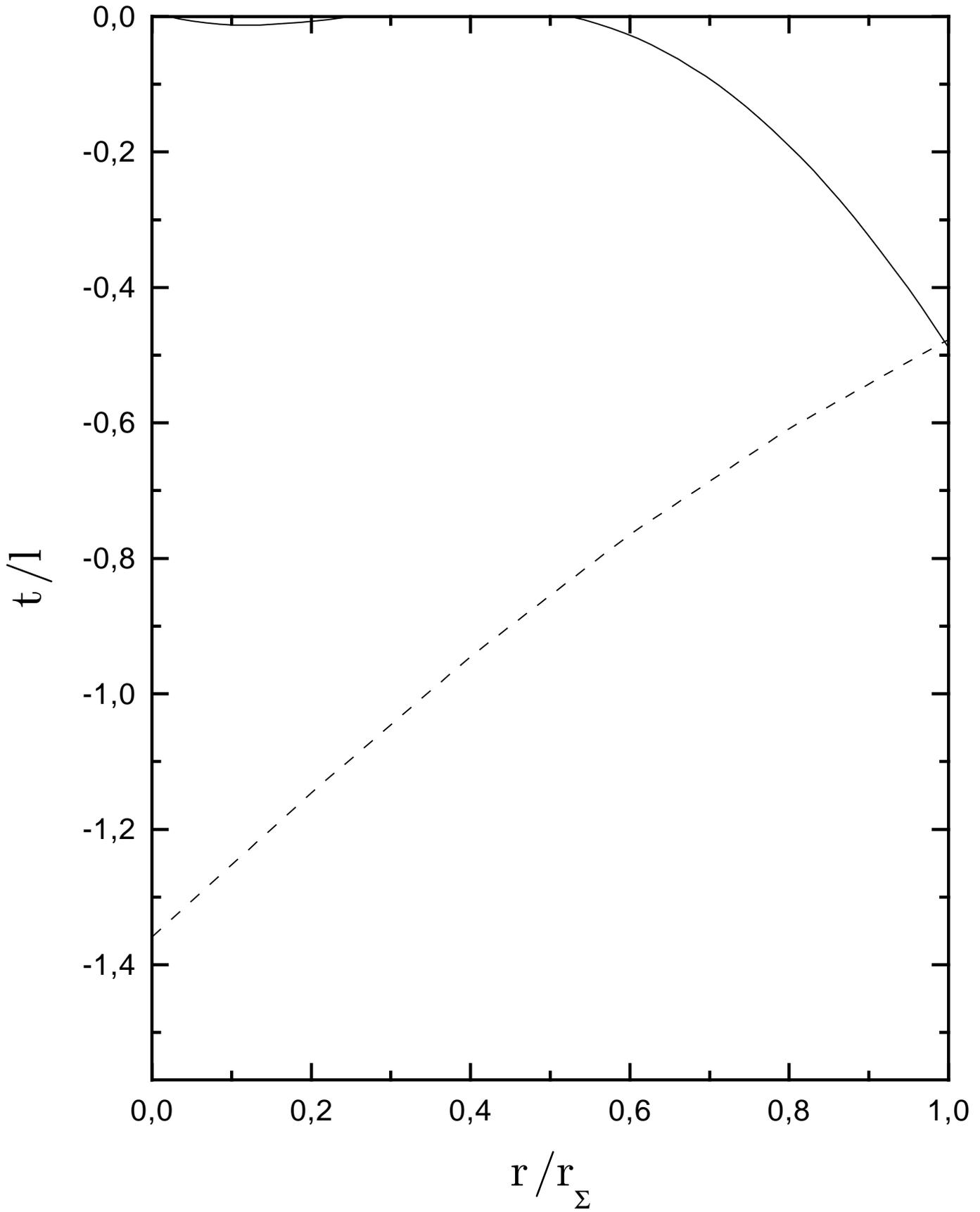
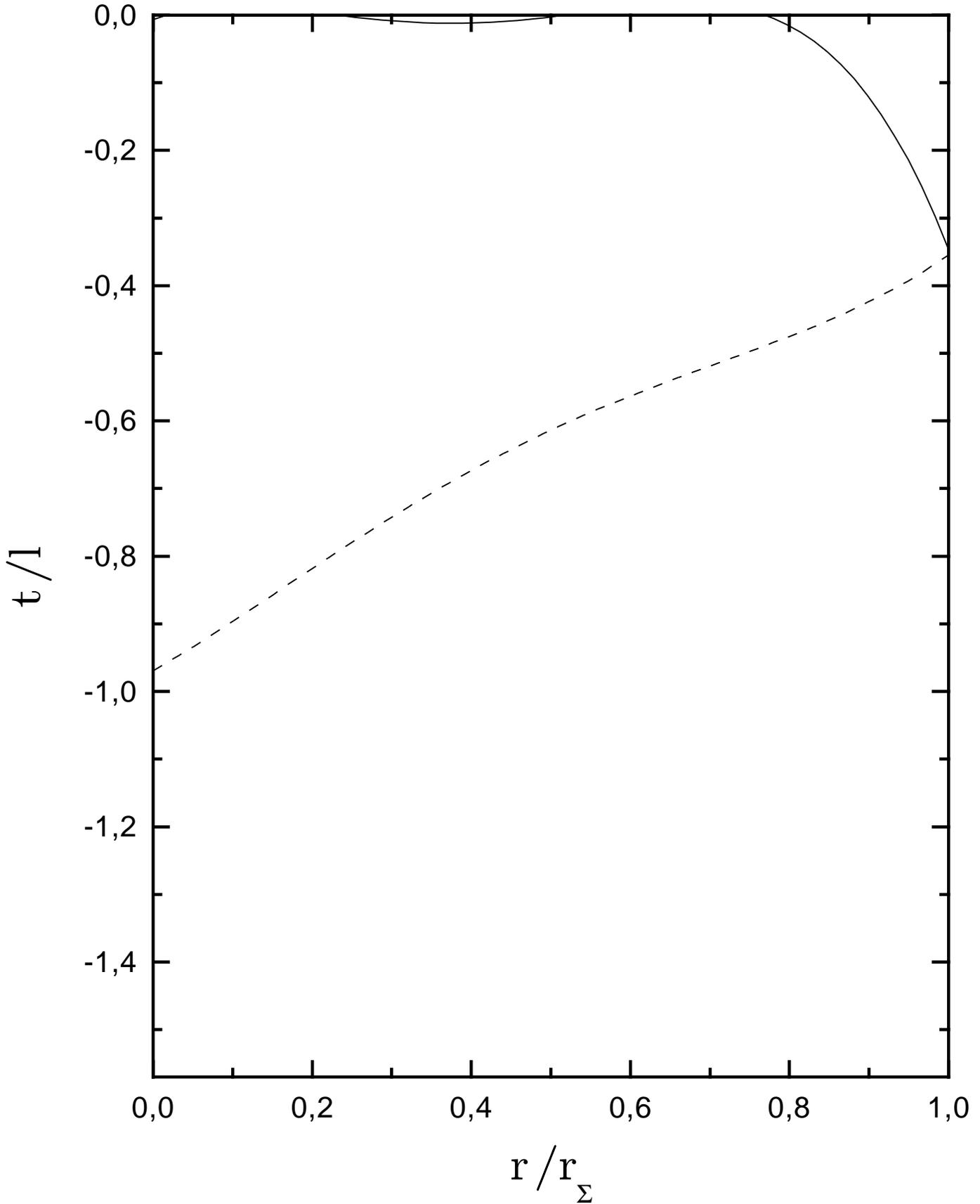
\end{document}